\documentclass[showpacs,aps,prd,nofootinbib,floatfix,amsmath,amssymb]{revtex4}
\usepackage{graphicx}
\usepackage{multirow}
\newcommand{\uu}{{\mathcal{U}}}
\newcommand{\du}{{d_\mathcal{U}}}
\newcommand{\dec}{\ell_i^-\to\ell_j^-\ell_k^- \ell_k^+}

\begin{document}

\title{Lepton electric and magnetic dipole moments via lepton flavor
violating spin-1 unparticle interactions}

\author{ A. Moyotl}
\affiliation{Instituto de F\'isica,  Benem\'erita Universidad
Aut\'onoma de Puebla, Apartado Postal 72570 Puebla, M\'exico}
\author{A. Rosado}
\affiliation{Instituto de F\'isica,  Benem\'erita Universidad
Aut\'onoma de Puebla, Apartado Postal 72570 Puebla, M\'exico}
\author{G. Tavares-Velasco}		
\email[E-mail:]{gtv@fcfm.buap.mx}
\affiliation{Facultad de
Ciencias F\'\i sico Matem\'aticas, Benem\'erita Universidad
Aut\'onoma de Puebla, Apartado Postal 1152, Puebla, Pue., M\'
exico}

\date{\today}
\begin{abstract}
The magnetic dipole moment (MDM) and the electric dipole moment (EDM) of
leptons are calculated under the assumption of lepton flavor violation (LFV)
induced by spin-1 unparticles with both vector and axial-vector couplings to leptons,
including a CP-violating phase. The experimental limits on the muon MDM and
 LFV process such as the decay $\dec$ are
then used to constrain the LFV couplings for particular values of the
unparticle operator dimension $\du$ and the unparticle scale $\Lambda_\uu$,
assuming that LFV transitions  between the tau and muon leptons are dominant. It
is found that the
current experimental constraints favor a scenario with dominance of the
vector
couplings over the axial-vector couplings. We also
obtain estimates for the EDMs of
the electron and the muon, which are well below the experimental values.
\end{abstract}

\date{\today}

\maketitle
\section{Introduction}

Scale invariant quantum field theories cannot interpret matter in terms
of particles. Motivated by the Banks and  Zaks model \cite{Banks:1981nn}, Georgi
\cite{Georgi:2007ek,Georgi:2007si} conjectured a scenario that, unlike the one
posed by the SM and  its extensions, introduces scale invariant
stuff associated with fractionary anomalous dimension operators. Georgi
suggested that a yet unseen scale invariant sector may exist in the high energy
theory such that scale invariant stuff may interact weakly with the SM fields.
Such a hidden sector may manifest itself at an energy scale $\Lambda_{\uu} > 1$
TeV, and since physical particles cannot exist in this sector, the interactions
with the SM fields would occur through scale invariant fields known as
unparticles. Although a comprehensive study of this class of theories is very complex,
it is possible to describe its low energy effects
through an effective field theory. This allows one to study the
phenomenological effects of unparticle stuff.

The  appropriate theoretical framework to describe unparticle physics is the one
introduced by Banks and  Zaks \cite{Banks:1981nn}.  The hidden sector is a
${\mathcal B}{\mathcal Z}$ sector and the associated fields  are described by
renormalizable ${\mathcal O}_{{\mathcal B}{\mathcal Z}}$ operators. It is
assumed that these fields interact with the SM fields through the exchange of
heavy particles at a very high energy ${\mathcal M}_{\uu}$. Below this energy
scale there is nonrenormalizable couplings between the fields of the ${\mathcal
B}{\mathcal Z}$ sector and the SM fields,\footnote{The dimension of the
respective operators are $d_{{\mathcal B}{\mathcal Z}}$ and $d_{SM}$} which
generically can be written as $ {\mathcal O}_{SM}{\mathcal O}_{{\mathcal
B}{\mathcal Z}}/{\mathcal M}_{\uu}^{d_{SM}+d_{{\mathcal B}{\mathcal Z}}-4}$.
Dimensional transmutation is caused by the renormalizable couplings of the
${\mathcal B}{\mathcal Z}$ sector at an energy scale $\Lambda_{\uu}$ as scale
invariance emerges. Below this energy scale, an effective theory can be used to
describe the interactions between the SM fields and the ${\mathcal B}{\mathcal
Z}$ fields, which are associated with unparticles. The effective Lagrangian can
be written as \cite{Georgi:2007ek,Georgi:2007si}:

\begin{equation}
{\mathcal L}_{\uu}=C_{{\mathcal O}_{\uu}} \frac{\Lambda_{\uu}^{d_{{\mathcal B}{\mathcal Z}}-\du}}{{\mathcal M}_{\uu}^{d_{SM}+d_{{\mathcal B}{\mathcal Z}}-4}} {\mathcal O}_{SM}{\mathcal O}_{\uu}, \label{efflag}
\end{equation}

\noindent where $C_{{\mathcal O}_{\uu}}$ stands for the coupling constant and
the operator dimension $\du$ can be fractionary. From theoretical considerations
\cite{Mack:1975je}, it has been noted \cite{Nakayama:2007qu,Grinstein:2008qk}
that unitarity is guaranteed in the interval $\du>1$. The Lorentz structure of
unparticle operators is nontrivial but it can be constructed from the nature of
the primary operator  ${\mathcal O}_{{\mathcal B}{\mathcal Z}}$ and its
transmutation. Such a Lorentz structure can be scalar, ${\mathcal O}_{\uu}$,
vector, ${\mathcal O}_{\uu}^\mu$, spinor or tensor, ${\mathcal
O}_{\uu}^{\mu\nu} $. The respective propagators  of these unparticles
along with  their interactions with the SM particles  have been studied with detail in
\cite{Georgi:2007ek,Banks:1981nn,Luo:2007bq,Chen:2007qr,Liao:2008tj,
Cheung:2007zza}.

Shortly afterwards the unparticle idea came out, the study of its phenomenology
was eagerly addressed \cite{Luo:2007bq,Cheung:2007zza,Cheung:2007ap}. One
interesting effect could arise from the interference between unparticle and SM
amplitudes, such as could occur in the Drell-Yan process
at the Tevatron and the LHC
\cite{Cheung:2007zza,Mathews:2007hr}: in particular, the most peculiar effects
could be observed in the invariant dilepton invariant mass distribution near the
$Z$-pole \cite{Cheung:2007zza,Mathews:2007hr}. This class of interference
effects could also be evident in  diphoton  production at the LHC
\cite{Kumar:2007af}. As far as the direct production of unparticles is
concerned, it has been studied through mono-photon, $e^-e^+\to \gamma\uu$, and
mono-$Z$ production, $e^-e^+\to Z\uu$ \cite{Cheung:2007ap}, whereas  the
production of an unparticle accompanied by a mono-jet was studied in
\cite{Rizzo:2008fp}.  Several decays of SM particles into unparticles have been
examined: $t\to b\uu$ \cite{Georgi:2007ek}, $Z\to \bar{f}f\uu$
\cite{Cheung:2007ap,Cheung:2007zza}, $H\to \gamma\uu$ \cite{Cheung:2007sc}, and
$Z\to \gamma\uu$ \cite{Cheung:2008ii}.  In addition, other topics on unparticle
physics have been studied, such as the possible effects of unparticles on
CP violation \cite{Chen:2007vv,Huang:2007ax}, lepton flavor violation (LFV)
\cite{Aliev:2007qw,Lu:2007mx,Ding:2008zza}, and lepton electromagnetic
properties \cite{Liao:2007bx,Hektor:2008xu,Iltan:2007ve}.

In order to impose constraints on unparticle parameters, several experimental
data have been used.  The $e^-e^+\to \gamma\uu$ process was studied to explain
$\gamma\bar\nu\nu$ production at LEP \cite{Cheung:2007ap}. It was found that LEP
data are consistent with the values  $\Lambda_{\uu} =1.35$ TeV  for $\du=2$ and
$\Lambda_{\uu} =660$ TeV for $\du=$1.4. Unparticle constraints have also been
obtained from experimental data on cosmology and astrophysics
\cite{Davoudiasl:2007jr,Hannestad:2007ys,Das:2007nu,Freitas:2007ip}.

Apart from the tree-level effects, loop induced unparticle effects have studied
in the literature
\cite{Liao:2007bx,Hektor:2008xu,Ding:2008zza,Iltan:2008jn}. The
electron magnetic dipole moment via scalar and vector unparticles was obtained
in \cite{Luo:2007bq,Liao:2007bx}. This study was later extended  for the muon
magnetic moment due to scalar unparticles with LFV couplings
\cite{Hektor:2008xu}, whereas the lepton electric dipole moment via scalar unparticles was studied in
\cite{Iltan:2007ve}. In addition, the calculation of the fermion dipole moments via fermion
unparticles was presented in \cite{Liao:2008tj}.  The study of loop induced
decays mediated by unparticles has also been addressed, for instance the decays  $l_i\to l_j\gamma$
\cite{Hektor:2008xu,Ding:2008zza} and $Z\to \bar{l}_i l_j$
\cite{Iltan:2008jn}. In this work we are interested in calculating the spin-1
unparticle contribution to the magnetic dipole moment (MDM) and the electric dipole
moment (EDM) of leptons in the most general case when there is LFV interactions.
To our knowledge
this calculation has not been presented in the literature.

The rest of the work is organized as follows. In Sec. II we present an overview
of unparticle operators. Section III is devoted to the results for the lepton
electromagnetic vertex mediated by vector  unparticles, while the
numerical analysis is presented in Sec. IV. The conclusions and outlook are
presented in Sec. V.

\section{Unparticles interactions with the SM fields}

The interactions between the SM particles and unparticles occur through the
exchange of heavy fields of mass ${\mathcal M}_{\uu}$. Once those heavy fields
are integrated out, the effective Lagrangian that describes the interactions
between the SM particles and unparticles is obtained.  This effective Lagrangian
is composed by  a tower of effective operators that can be constructed out of
the SM fields by invoking the $SU_L(2)\times U_Y(1)$ gauge symmetry.
For instance, the effective interactions of spin-0 and
spin-1 unparticles with SM fermions are as follows
\cite{Georgi:2007ek,Banks:1981nn,Luo:2007bq,Chen:2007qr,Cheung:2007zza}:

\begin{eqnarray}
{\mathcal L}_{\uu_S}&=&  \frac{\lambda_{S}^{ij}}{\Lambda_{\uu}^{\du-1}} \bar{f}_{i} f_{j} {\mathcal O}_{\uu} +\frac{\lambda_{P}^{ij}}{\Lambda_{\uu}^{\du-1}} \bar{f}_i \gamma^5 f_j {\mathcal O}_{\uu},
                                            \label{scaint}\\
{\mathcal L}_{\uu_V}&=  &\frac{\lambda_{V}^{ij}}{\Lambda_{\uu}^{\du-1}}
\bar{f}_i\gamma_\mu f_j {\mathcal O}_{\uu}^\mu
+\frac{\lambda_{A}^{ij}}{\Lambda_{\uu}^{\du-1}} \bar{f}_i\gamma_\mu  \gamma^5 f_j
{\mathcal O}_{\uu}^\mu.
                                            \label{vecint}
\end{eqnarray}

\noindent  where $i$ and $j$ stand for the family index and
$\lambda^{ij}_J=C_{{\mathcal O}_{\uu}} \Lambda_{\uu}^{d_{{\mathcal B}{\mathcal Z}}}/{\mathcal M}_{\uu}^{d_{SM}+d_{{\mathcal B}{\mathcal Z}}-4}$ stands for the associated coupling constants. These effective operators break scale invariance.

Due to the invariant scale nature of unparticles, their propagators  can be
constructed by means of unitary cuts and the spectral decomposition formula.
Therefore, the scalar unparticle propagator is given by:

\begin{equation}
\Delta_F(p^2)= \frac{A_{\du}}{2\sin(\du \pi)} (-p^2-i\epsilon)^ {\du-2}
\end{equation}

\noindent where the $A_{\du}$ function, which is introduced to normalize the
spectral density \cite{Cheung:2007ap}, is given as follows:

\begin{equation}
A_{\du}=\frac{16\pi^2\sqrt{\pi}}{(2\pi)^{2\du}}\frac{\Gamma(\du+\frac{1}{2})}{\Gamma(\du-1)\Gamma(2\du)}
\end{equation}

\noindent As far as the vector unparticle is concerned, its propagator is given
by

\begin{eqnarray}
\Delta_F^{\mu\nu}(p^2)& = &\Delta_F(p^2) \pi^{\mu\nu}(p),
\end{eqnarray}
where $\pi^{\mu\nu}(p)$ is given explicitly as
\begin{eqnarray}
\pi^{\mu\nu}(p)&=&-g^{\mu\nu}+a  \frac{p^\mu p^\nu}{p^2}.
\end{eqnarray}
The form of this expression is due to the spin structure of this class of
unparticles \cite{Cheung:2007ap}. Also $p^2\ne M^2$, which is a reflect of the
unphysical nature of unparticles. When the unparticle field is taken as
transverse, it turns out that
$p_\mu\pi^{\mu\nu}(p)=0$,
which translates into the condition $a=1$. However, in the context of a specific conformal invariance, $a=2(\du-2)/(\du-1)$ \cite{Grinstein:2008qk}. In the limit $\du \to 1^+$
the propagator $\Delta_F(p^2)$ turns out to be the propagator of a massless
scalar particle, as expected:

 \begin{equation}
 \lim_{\du \to 1^+}\Delta_F(p^2)=\frac{1}{p^2}.
 \end{equation}

\section{Unparticle contribution to electric and magnetic dipole moments of
leptons}

Among the best measured particle observables,  the muon MDM, $a_\mu$, stand outs: it has been measured with an
impressive accuracy of
$0.54$ ppm. The current world average, which is dominated by the measurements of
the E281 collaboration at BNL, is given by \cite{Bennett:2006fi}
\begin{equation}
a_\mu^{\rm Exp.}=116592089\,\, (63)\times 10^{-11}\quad (0.54{\rm ppm}),
\end{equation}
where the statistical and  systematic errors, $0.46$ ppm and  $0.28$
ppm, have been added in quadrature.

The SM theoretical prediction is given by the sum of the QED,
electroweak and the hadronic contributions. A recent review of these
calculations is given in \cite{Passera:2004bj}. While the QED and electroweak
contributions to $a_\mu$ have been calculated with a great precision, the
largest uncertainty arises from the hadronic contribution, which is still under
revision. The theoretical SM prediction is
\begin{equation}
a_\mu^{\rm SM}= 116591834\,\,(48)\times 10^{-11}.
\end{equation}
where the $\sigma(e^- e^+\to {\rm hadrons})$ data have been used to calculate
the leading-order hadronic vacuum polarization contribution
\cite{Davier:2009zi}. There is thus a  discrepancy  between the
experimental and theoretical predictions larger than 3.6 standard deviations
\cite{Nakamura:2010zzi}:
\begin{equation}
\Delta{a_\mu}=a_\mu^{\rm Exp.}-a_\mu^{\rm SM}=255\,(80)\times 10^{-11}.
\end{equation}
As long as this disagreement is attributed entirely to
new physics \cite{Jegerlehner:2009ry}, the allowed minimal an maximal limits for these
contributions, with 95\% C.L are $\Delta a_\mu =(255\mp 1.96\times 80) \times
10^{-11}$.  It may be however that such a
discrepancy will reduce to an acceptable level once the hadronic contribution is
determined with best accuracy.

Another well studied fermion electromagnetic property  is the EDM, $d_f$, which can provide an excellent probe of new sources of
CP-violation. In the SM, the Cabbibo-Kobayashi-Maskawa (CKM) mechanism  cannot
account for the amount of  CP violation required to explain the baryogenesis
asymmetry of the universe. It has been long known that the experimental
observation of an EDM of fundamental particles will hint to
new physics as the SM predictions are very small. For instance,  the electron EDM is predicted to be  negligibly small as it arises
up to the three-loop level via the CKM phase. Other models such as supersymetry,
multi Higgs and the left-right symmetric model predict much larger values for
$d_e$ \cite{Bernreuther:1990jx}. The current experimental limit on the electron
EDM with 90\% C.L. is  \cite{Regan:2002ta}:

\begin{equation}
 |d_e|\le 1.6\times 10^{-27}\,\,{\rm e \,cm},
\end{equation}
whereas the experimental limits on the positive and negative muon
EDMs  with 95\% C.L. are
\cite{Bennett:2008dy}:
\begin{eqnarray}
 |d_{\mu}^+|&\le&2.1\times 10^{-19}\,\,{\rm e \,cm},\\
 |d_{\mu}^-|&\le&1.5\times 10^{-19}\,\,{\rm e \,cm}.
\end{eqnarray}

We will calculate the contribution
to the lepton MDM and EDM induced by a spin-1 unparticle with both vector and axial-vector  LFV
couplings.

\subsection{Lepton dipole moments via spin-1 unparticle}

\begin{figure}[!ht]
\includegraphics[width=7cm]{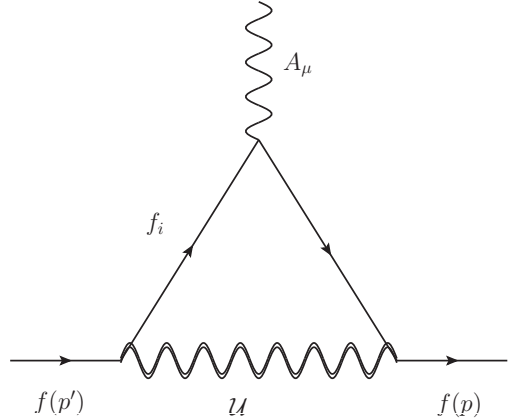}
\caption{Feynman diagram for the lepton electromagnetic vertex induced by
unparticles.\label{vertex}}
\end{figure}

The Feynman diagram contributing to the electromagnetic vertex is shown in Fig.
\ref{vertex}. From Eqs. (\ref{scaint}) and (\ref{vecint}) we obtain the interactions of the spin-1
unparticle with both vector and axial-vector LFV  couplings. We will use Feynman parameters for the calculation and the
spectral form for the unparticle propagator will be considered:
\begin{equation}
\Delta_F(p^2)=\frac{A_{\du}}{2\pi}\int_0^\infty
\frac{dm^2\,(m^2)^{\du-2}}{p^2-m^2+i\epsilon},
\end{equation}
After the momentum space integration is worked out, we will proceed with the
spectral integral.

After some lengthy  algebra we arrive at the following
results. The MDM of the lepton $j$ due to spin-1
unparticle can be written as
\begin{eqnarray}
 a_j^\uu&=&\sum_{J=V,A}\,
\sum_{i=e,\mu,\tau} |\lambda_{J}^{ij}|^2
F_{J}(m_i,{\du}),
\label{amu}
\end{eqnarray}
where the $F_J$ functions can be written as
\begin{equation}
F_J(m_i,\du)= \frac{A_{\du}}{16 \pi^2\sin(\pi\du)}
\left(\frac{m_i^2}{\Lambda^2_\uu}\right)^{\du-1}f_J\left(\sqrt{r_i},
\du\right),
\label{FJ}
\end{equation}
with $\sqrt{r_i}=m_j/m_i$ and
\begin{eqnarray}
f_{V}(z,{\du})&=& \frac{-z}{2-\du}\int^1_0 dx\,\left(1-x\right)^{\du-1}x^{2-\du}
(1-z^2x)^{\du-3} \Big(
(3(3-x)-4 \du)+z \left(3 (\du-1) x+\du-3\right)\nonumber\\&+&z^2\left( (1+x)^2\du+(x-5) x-2\right)-z^3\left(3 (\du-1)
x+\du-3\right)x\Big),
\end{eqnarray}
also $f_{A}(z,{\du})=f_{V}(-z,{\du})$. We note  that there is a flip
in the sign of the vector  and axial-vector couplings.

For our analysis below, we will also need the contribution to $a_\mu$ from a spin-0
unparticle with both scalar and pseudoscalar LFV couplings. We obtain a similar result as that given by Eqs.
(\ref{amu}) and (\ref{FJ}), with $J$ running over $S$ and $P$, while the $f_S$
function is

\begin{equation}
f_{S}(z,{\du})= -z \int^1_0 dx\, \left(1-x\right)^{\du}x^{1-\du}(1-z^2
x)^{\du-2} \left(1+zx\right),
\end{equation}
and $f_{P}(z,{\du})=f_{S}(-z,{\du})$. These results coincide with those presented in \cite{Hektor:2008xu}
and serve as a cross-check for our calculation method.

We would also like to note that
in the case of diagonal unparticle couplings, we obtain for the spin-0 and spin-1 unparticle contributions to $a_\ell$:

\begin{eqnarray}
a_\ell^\uu&=&\frac{A_{\du}\Gamma (2-\du) \Gamma (2
\du-1)}{16\pi^2\sin(\pi\du)\Gamma (\du+2)}
\left(\frac{m_f^2}{\Lambda^2_\uu}\right)^{\du-1}\left(2
(\du-2)|\lambda_{V}^{\ell}|^2  +\frac{4(2- \du)}{\du-1}|\lambda_{A}^{\ell}|^2
-3|\lambda_{S}^{\ell}|^2+(2\du-1)|\lambda_{P}^{\ell}|^2\right),
\end{eqnarray}
where $\lambda^\ell_J=\lambda_J^{\ell\ell}$ stands for the diagonal unparticle couplings. This result agrees with the result presented in \cite{Liao:2007bx} for the electron
MDM.

As far as the lepton EDM is concerned, we will consider the contribution from both spin-0 and spin-1 unparticles. The EDM of fermion $j$, which can only arise if both
$\lambda_V^{ij}$ and $\lambda_A^{ij}$  ($\lambda_S^{ij}$ and $\lambda_P^{ij}$)
are nonzero and have an imaginary phase, is given by

\begin{eqnarray}
 d_j^\uu&=&\sum_{(J,K)}\sum_{i=e,\mu,\tau}{\rm
Im}\left(\lambda_{J}^{ij} {\lambda_K^{*ij}}\right)G_{(J,K)}(m_i,{\du}),
\label{dj}
\end{eqnarray}
where  $(J,K)$ runs over $(V,A)$ and $(S,P)$. The $G_{(J,K)}$
functions are defined as
\begin{equation}
G_{(J,K)}(m_i,\du)=\frac{e\,A_{\du}}{32
\pi^2\sin(\pi\du)\, m_i}
 \left(\frac{m_i^2}{\Lambda^2_\uu}\right)^{\du-1}g_{(J,K)}\left(\sqrt{r_i},
\du\right)
\end{equation}
with
\begin{eqnarray}
g_{(V,A)}(z,{\du})&=&\frac{1}{2-\du}\int_0^1 dx\,
\left(1-x\right)^{\du-1}x^{2-\du} (1-z^2x)^{\du-3} \Big(4 (1-z^2x)(2-\du)+
(1-3 x)\nonumber\\&+&z^2\left((x^2-1)\du+(x-1)x+2\right)\Big),
\label{gVA}
\end{eqnarray}

\begin{eqnarray}
g_{(S,P)}(z,{\du})&=&- \int^1_0 dx\,
   \left(1-x\right)^{\du}x^{1-\du}
   (1-z^2x)^{\du-2} ,
\label{gSP}
\end{eqnarray}
In the limit of a heavy internal lepton, $m_j\ll m_i$, the integration can be
dealt with and we obtain
\begin{eqnarray}
G_{(V,A)}(m_i,\du)= \frac{3(\du-2) (\du-1) A_\du
  }{64 \pi \sin^2(\pi  \du) m_i} \left(\frac{m_i^2}{\Lambda_\uu^2 }\right)^{
\du-1},
\label{GVAsim}
\end{eqnarray}
\begin{eqnarray}
G_{(S,P)}(m_i,\du)= \frac{(\du-1) \du A_\du }
{64 \pi \sin^2(\pi  \du)  m_i}\left(\frac{m_i^2}{\Lambda_\uu^2}\right)^{
\du-1}.
\label{GSPsim}
\end{eqnarray}
These
expressions can be useful for the muon and tau loop contributions to the
electron
EDM and the tau loop contribution to the muon EDM.

\section{Numerical analysis and discussion}

Apart from the muon MDM, experimental limits on LFV process are known to be useful to constrain LFV couplings.
LFV processes involving the muon are among the most constrained by the
experiment. For instance, there are stringent constraints on LFV muon decays:
BR$(\mu \to e\gamma) < 2.4 \times 10^{-12}$ \cite{Adam:2011ch} and BR$(\mu\to
3e) < 1.0 \times
10^{-12}$ \cite{Bellgardt:1987du}. Furthermore, the bound on the $\mu \to
e\gamma$ rate is expected to be improved  by about one order of magnitude by the MEG
experiment \cite{Ritt:2006cg}. However, less stringent constraints  exist for LFV
$\tau$ transitions:  BR$(\tau\to \mu \gamma) < 4.4 \times
10^{-8}$ \cite{:2009tk}, BR$(\tau\to 3e) < 3.6 \times
10^{-8}$ \cite{Aubert:2007pw}, etc. Below we will analyze the constraints on LFV spin-1 unparticle couplings.

\subsection{Muon anomalous magnetic moment}

A comprehensive analysis of the spin-1 unparticle contribution to the muon MDM would require to deal with several free parameters: six coupling constants, the unparticle scale and the unparticle operator dimension.
We will take another approach instead and consider  some particular scenarios of LFV
together with the hypothesis that there is no large cancellation between
different unparticle contributions.
First of all, we will assume that there is a hierarchy in the LFV unparticle
couplings as it is assumed in other  models of LFV, i.e.,
$|\lambda^{e\mu}|< |\lambda^{e\tau }|< |\lambda^{\mu\tau}|\ll\lambda^{ii}$ for
all the
unparticle couplings. Since experimental data strongly
constrains LFV between the muon and the electron, we will concentrate
on the  scenario
where LFV transitions are dominated by the couplings of spin-1 unparticles with the tau and muon
leptons. More specifically, we will focus on the three following scenarios:

\begin{itemize}\itemsep1pt
\item[A-] Vector and axial-vector LFV couplings of similar size.
 \item[B-] Vector-dominated LFV couplings.
\item[C-]  Axial-vector-dominated LFV couplings.
\end{itemize}

\begin{figure}[!ht]
 \includegraphics[width=9cm]{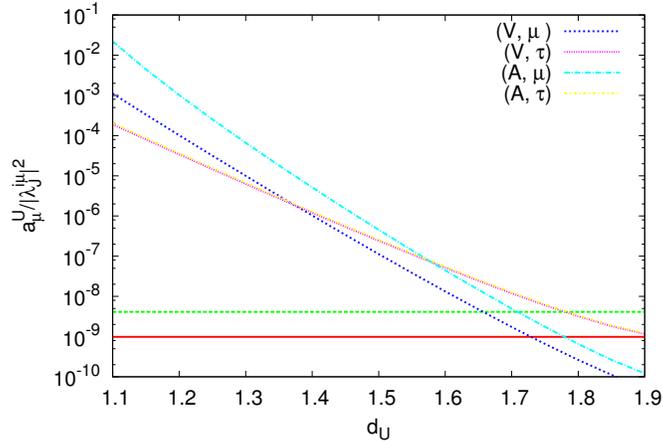}
\caption{Partial contributions from the indicated lepton loop  to the muon MDM
due to  vector $(V)$ or axial-vector $(A)$ unparticle couplings as a
function of the $\du$ dimension and for $\Lambda_\uu=1$ TeV. The absolute values
of the (negative) $(A)$ contributions are shown.
The horizontal lines are the minimal and maximal allowed limits on new physics contributions to $a_\mu$ with $95\%$
C.L. Notice that the contribution from the tau loop are almost indistinguishable.
\label{plot1}}
\end{figure}

\begin{figure}[!ht]
 \includegraphics[width=9cm]{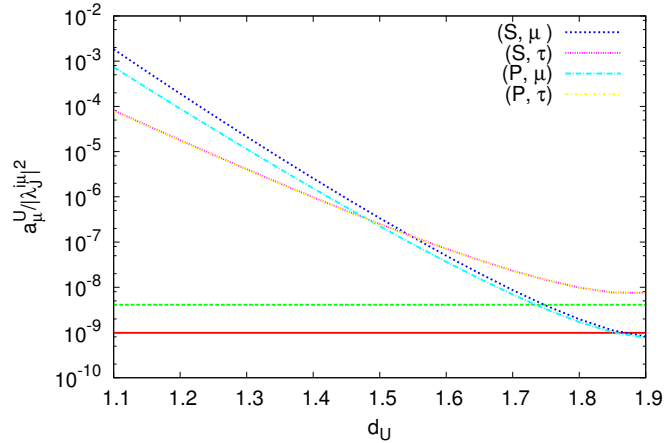}
\caption{The same as in Fig. \ref{plot1}, but for the scalar and pseudoscalar contributions to $a_\mu$. The absolute values
of the (negative) $(P)$ contributions are shown.
\label{plot2}}
\end{figure}

To discuss these scenarios we first evaluate numerically each term,
$F_J(m_i,\du)$, of the sum of Eq. (\ref{amu}). The results are shown in
Fig. \ref{plot1} as functions of the dimension $\du$ and for $\Lambda_\uu=1$
TeV.  We do not show the contributions due to the electron loop as we assumed that the LFV couplings involving the electron are subdominant. While the contributions from vector couplings are positive,
the axial-vector contributions are negative. We also note that $F_{A}(m_i,\du) \simeq -F_{V}(m_i,\du)$ and so the vector and axial-vector
contributions can largely cancel out. This effect is more evident for the
contributions of the tau lepton. Furthermore, the contributions from
axial-vector
couplings are larger in magnitude than the vector contributions, with the
largest contribution arising from the muon loop. It means that as long the
vector and axial-vector unparticle couplings were of similar order of magnitude
(scenario A)  the total contribution to the muon MDM could be negative. This is a scenario disfavored by the
current experimental data, which requires a positive contribution to $a_\mu^\uu$
from new physics. The situation is rather similar for the scalar and
pseudoscalar contributions, which for comparison purposes are shown in Fig. \ref{plot2}, although in
this case the scalar contributions are slightly larger in magnitude than the
axial-vector contributions. If all the unparticle couplings  were of similar
size, the unparticle contribution to $a_\mu^\uu$
could be negative as the axial-vector plus pseudoscalar contributions could be
dominant. However, even if  all the unparticle couplings were large, the total contribution to $a_\mu$ could
still be positive by a sort of fine-tuning.
We consider the scenario in which the
axial-vector  couplings are negligible in comparison to the vector
couplings (scenario B), such that the total contribution to $a_\mu^\uu$ is
positive. Below we will
analyze the possible constraints on vector unparticle couplings under this assumption.

\begin{figure}[!ht]
 \includegraphics[width=9cm]{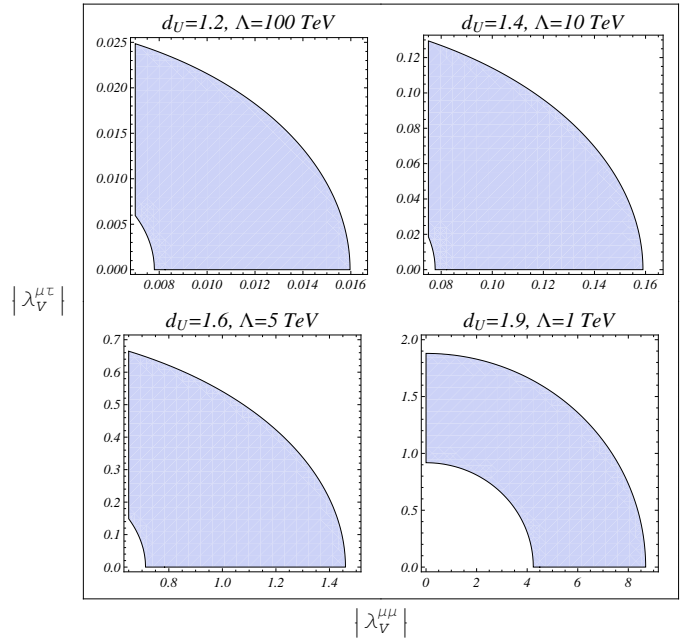}
\caption{Allowed area on the
$|\lambda^{\mu\tau}_V|$ vs $|\lambda^{\mu\mu}_V|$ plane consistent with the
experimental
limit on the muon MDM with $95 \%$ C.L.  We used the indicated values of
$\Lambda_\uu$  and  $\du$.
\label{plot3}}
\end{figure}

\begin{figure}[!ht]
 \includegraphics[width=9cm]{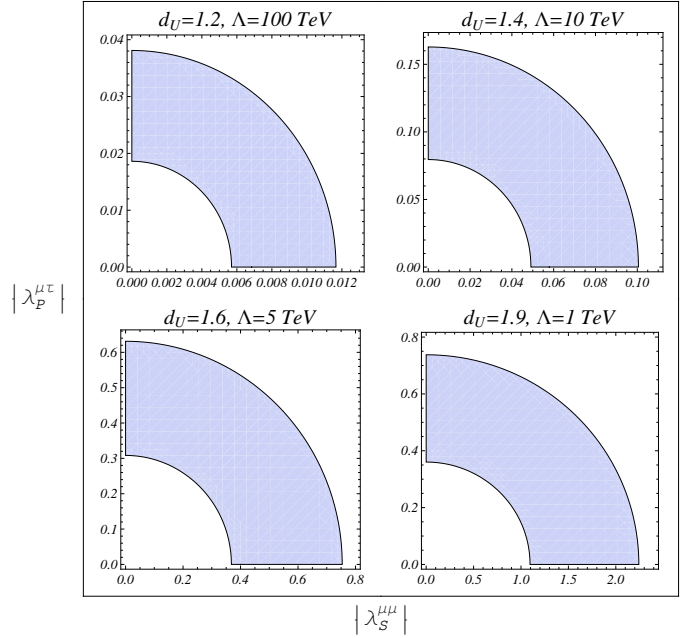}
\caption{The same as in Fig. \ref{plot3}, but for the bounds on the $|\lambda^{\tau\mu}_S|$ vs. $|\lambda^{\mu\mu}_P|$ plane.
\label{plot4}}
\end{figure}

If the current discrepancy between the theoretical SM
prediction and the experimental value of the muon MDM is assumed to be due
entirely to unparticle interactions, the allowed limits with 95 \% C.L. for this
class of contributions are $98.2\times 10^{-11}\le a_\mu^\uu\le 411.8 \times 10^{-11}$.
We have found the allowed area on the $|\lambda_V^{\tau\mu}|$
vs $|\lambda_V^{\mu\mu}|$ plane, which is shown in Fig.
\ref{plot3},
for several values of $\Lambda_\uu$ and $\du$ that are consistent with the
bounds obtained for the scale $\Lambda_\uu$  from mono-photon production at LEP
\cite{Cheung:2007ap} (see Table \ref{boundlam}). As inferred from
Fig.
\ref{plot3}, the experimental data allow a magnitude of the vector couplings as
large as unity for
$\du\simeq 2$ and $\Lambda_\uu=1$ TeV, whereas the most strong constraints are
obtained for $\du$
close to unity. In general the limits on $|\lambda^{\mu\tau}_V|$  are slightly stronger than the limits on
$|\lambda^{\mu\mu}_V|$, which is in agreement with our assumption. We also note that the bounds on the vector and axial-vector unparticle couplings are   of similar order of magnitude than the bounds on the scalar and pseudoscalar unparticle couplings, as shown in Fig. \ref{plot4}. To obtain that plot we have assumed that the dominant contribution to the muon MDM arise from a spin-0 unparticle.

\begin{table}[!ht]
\begin{tabular}{cc}
\hline
\\
$\du$& $\Lambda_\uu$ [TeV]\\
\hline
2.0&1.35\\
1.8&4\\
1.6&23\\
1.4&660\\
\hline
\end{tabular}
\caption{Constraints on $\Lambda_\uu$ from mono-photon production data
at LEP with $95\% $C.L. and assuming $\lambda_V^{ee}=1$
\cite{Cheung:2007ap}. \label{boundlam}}
\end{table}

\subsection{Bounds from the decay $\tau\to 3\mu$}
It has been long known that LFV couplings can also be constrained from the
experimental bounds on the tree-level induced decays
$\dec$. We will examine the bounds obtained  from
the $\tau\to 3\mu$
decay, which involves the $\lambda_V^{\mu\mu}$ and $\lambda_V^{\tau\mu}$
couplings. The calculation for the $\dec$ decay
width was already presented
in \cite{Aliev:2007qw} for the scalar unparticle contribution. We have
calculated the contribution from vector unparticles
and the result is presented in the Appendix \ref{appendix}.
Let us consider the scenario we are working in, and  neglect the axial-vector contributions. With
these assumptions we can write for the $\tau\to 3 \mu$ branching ratio:

\begin{eqnarray}
{\rm BR}(\tau\to 3\mu) &=&\frac{m_\tau \tau_\tau}{2^8
\pi^3}\left|\frac{A_\du}{\sin (d
\pi)}\right|^2
\left(\frac{m_\tau}{\Lambda_\uu}\right)^{4 (\du -1)}
|\lambda_V^{\mu\mu}|^2 |\lambda_V^{\mu\tau}|^2
\eta_1\left(\frac{m_\mu}{m_\tau},\du\right)\lesssim 10^{-8},
\label{taudecay}
\end{eqnarray}
with $\tau_\tau$ the tau mean life. The right-hand side of the inequality is the
experimental constraint  \cite{Nakamura:2010zzi} and the $\eta_1$ function is
defined in Appendix \ref{appendix}.

\begin{figure}[!ht]
 \includegraphics[width=9cm]{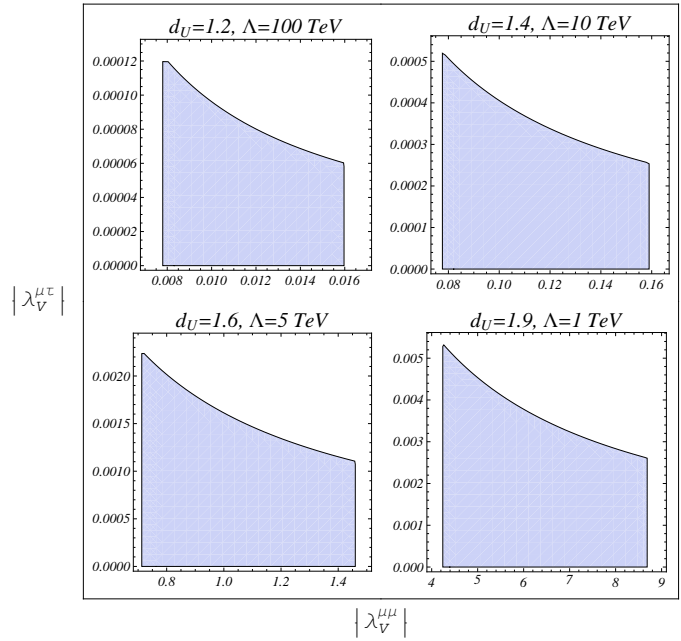}
\caption{Allowed area on the
$|\lambda^{\mu\tau}_V|$  vs $|\lambda^{\mu\mu}_V|$ plane consistent with  the
experimental
constraints on the muon MDM and the $\tau\to 3\mu$ decay for the
values of $\Lambda_\uu$  and  $\du$ indicated in each plot, when the dominant
contribution to these observables arise from the vector unparticle couplings..
\label{plot5}}
\end{figure}

\begin{figure}[!ht]
 \includegraphics[width=9cm]{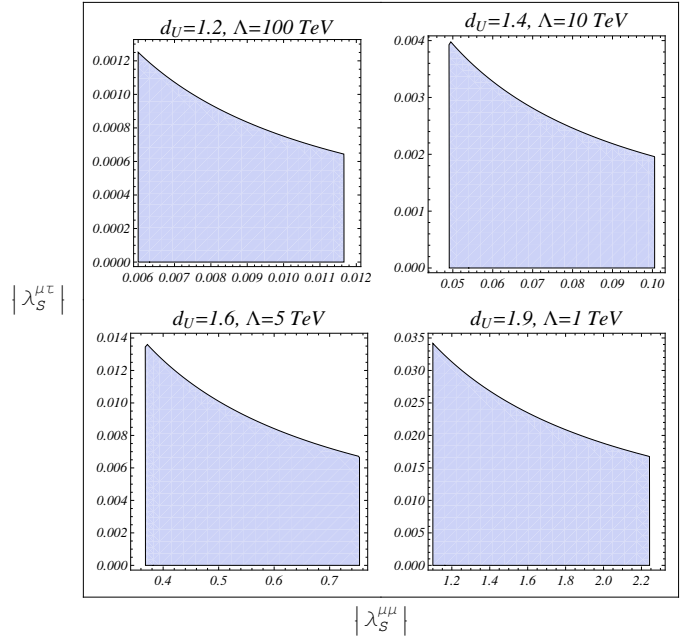}
\caption{The same as in Fig. \ref{plot5}, but when the dominant
contribution to $a_\mu$ and $\tau \to 3\mu$ arise from the scalar unparticle couplings.
\label{plot5s}}
\end{figure}

Eqs. (\ref{amu}) and (\ref{taudecay}) can serve to further constrain the
allowed values of the coupling constants. We show in Fig. \ref{plot5} the
allowed area obtained after numerical evaluation of Eq.
(\ref{taudecay}) for several values of $\Lambda_\uu$ and $\du$.
We observe that the $\tau\to3\mu$ decay constrains
considerably the size of the $\lambda_V^{\mu\tau}$ parameter, whose allowed
value is extremely small for $\du$ close to unity, where the
$\lambda_V^{\mu\mu}$ allowed values are considerably  larger. There is also an
allowed area in which the opposite is true ($|\lambda_V^{\mu\tau}|\gg
|\lambda_V^{\mu\mu}|$), but we will not consider it as conflicts with our
previous assumptions. For comparison purposes, we also have obtained the allowed area on the
$|\lambda^{\mu\tau}_S|$  vs $|\lambda^{\mu\mu}_S|$ plane when it is considered
that
LFV scalar unparticle couplings give the main contributions to the muon MDM
and the $\tau\to 3 \mu$ decay. The results are shown in Fig.
\ref{plot5s} for several values of $\Lambda_\uu$ and $\du$. We observe that the
allowed size of the scalar unparticle couplings is of similar order of magnitude
than for the vector unparticle couplings.

We also analyzed the possible constraints obtained from the $\tau\to\mu\gamma$
decay. It involves the $\lambda_V^{\mu\tau}$, $\lambda_V^{\mu\mu}$ and
$\lambda_V^{e\tau}$ couplings. We find that the respective constraints are less
stringent than the ones obtained in Fig. \ref{plot5}, so we will not
consider this decay mode.

\subsection{Electron and muon electric dipole moment}

\begin{figure}[!ht]
 \includegraphics[width=10cm]{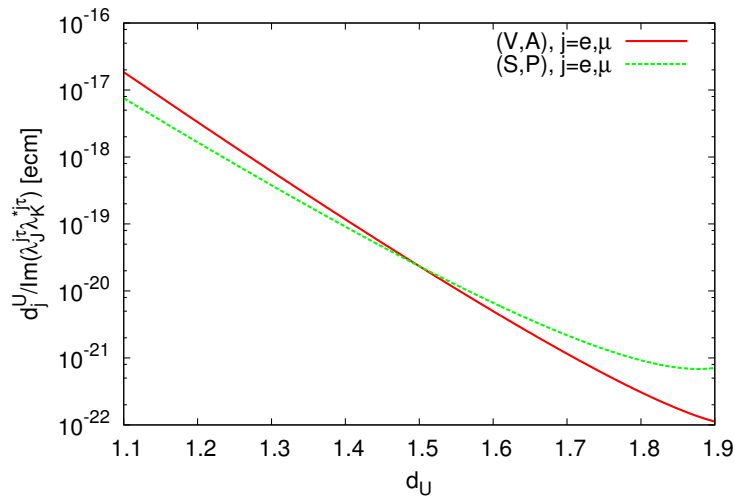}
\caption{Partial contribution from  the
tau lepton loop to the electron and muon EDM
due to vector-axial $(V,A)$ and scalar-pseudoscalar $(S,P)$ unparticle couplings
as a function of the $\du$ dimension and for $\Lambda_\uu=1$ TeV. The absolute
value of the (negative) $(V,A)$ contributions is shown. The curves for the tau contribution to the
electron and muon EDMs are indistinguishable.
\label{plotde}}
\end{figure}

In order to have a nonzero EDM,  both kind of unparticle
couplings, vector-axial or scalar-pseudoscalar, must be nonzero. Also, an
imaginary phase is required.
We will try to get an estimate of the order of magnitude for
the electron and muon EDMs induced by unparticles. Apart from the magnitude of
the $|\lambda_{V,A}^{ij}|$ and $|\lambda_{S,P}^{ij}|$ parameters, the EDM
depends on the associated CP-violating phases, so its analysis is even more
complicated. The most general form for the
unparticle couplings is $\lambda_J^{ij}=|\lambda_J^{ij}|\exp(-i\theta_J^{ij})$,
where $\theta_J^{ij}=-\theta_J^{ji}$ is a CP-violating phase. Therefore ${\rm
Im}(\lambda_{J}^{ij} {\lambda_K^{*ij}})=|\lambda_{J}^{ij}|
|\lambda_K^{ij}|\sin\delta\theta_{(J,K)}^{ij}$, with
$\delta\theta_{(J,K)}^{ij}=\theta_J^{ij}-\theta_K^{ij}$ the relative phase
between $J$ and $K$ couplings. Depending on the relative sign of the CP
violating phases,
$\delta\theta_{(V,A)}^{ij}$ and $\delta\theta_{(S,P)}^{ij}$, the partial
contributions can add coherently or destructively. We will
content ourselves with considering a somewhat restrictive scenario, which will
allows us to get an estimate of the order of magnitude of the EDM.
First of all, as was the case for the muon MDM, we will assume that there
is no large cancellation between different contributions to the EDM, so we
can analyze each contribution separately. Also, to be consistent with the
previous discussion, we will assume the following hierarchy
$|\lambda_{V,S}^{\mu\tau}|\gg |\lambda_ {V,S}^{\mu e}|$ and
$|\lambda_{V,S}^{e\tau}|\gg |\lambda_{V,S}^{e\mu}|$. It means
that the EDM would be entirely driven by the tau loop contributions. In view
of this scenario, we numerically evaluate the tau loop contributions to the
$G_{(J,K)}$ functions of Eq. (\ref{dj}). The
results are shown in Fig. \ref{plotde}.  We observe that the contribution from the tau loop to the electron and muon EDMs are
indistinguishable, which means that the EDM is entirely controlled by
the magnitude of the couplings constants and the imaginary phases.
For this calculation we used the approximate formulas given by Eqs. (\ref{GVAsim}) and
(\ref{GSPsim}), and made a cross-check with the exact calculation obtained
by numerical integration  of Eqs.
(\ref{gVA}) and (\ref{gSP}).

We now consider the constraints on the coupling constants obtained above to
estimate the muon EDM in our working scenario. Both $(V,A)$ and $(S,P)$
contributions are of the order of $10^{-17}$ e-cm for $\du\simeq 1.1$ and
$10̣^{-21}$ e-cm
for $\du\simeq 1.6$. We will consider the case in which the $(V,A)$ coupling
prevails over the $(S,P)$ couplings. The allowed magnitude of the
$|\lambda_V^{\mu\tau}|$ parameter is of the order of $10^{-5}$ for $\du=1.1$ and
$10^{-3}$ for $\du=1.6$. A good assumption for the $|\lambda_A^{\mu\tau}|$
coupling is that its magnitude  is about one or two orders of
magnitude below $|\lambda_V^{\mu\tau}|$. This  would not  spoil the
bounds we obtained above. Therefore the muon EDM is about
$|d_\mu^\uu|\simeq |\sin\delta_{(V,A)}^{\mu\tau}|\times 10^{-30}$ e-cm for
$\du\simeq 1.1$,
whereas $|d_\mu^\uu|\simeq |\sin\delta_{(V,A)}^{\mu\tau}|\times 10^{-29}$ e-cm
for
$\du\simeq
1.6$. These results are well below from the experimental limit on the muon EDM
as
the imaginary phase
is expected to be very small. The magnitude of the
coupling
constants for LFV transitions involving the electron are expected to be more
suppressed as long as the hierarchy discussed above is respected. Then we  can
estimate that the electron EDM induced by vector unparticles is well below the
$10^{-30}$ e-cm level and far from the experimental measurement. Moreover, since the constraints on LFV spin-0 and spin-1 unparticle couplings are similar, the contributions to the muon and electron EDMs from both spin-0 and spin-1 unparticles are expected to be of similar size. Even if all
different unparticle contributions add coherently, it is hard to expect a large
size of the electron and muon EDMs due to unparticles.

\section{Concluding remarks}
The lepton magnetic and electric dipole moments were calculated in the
framework of unparticle physics assuming interactions mainly driven by a spin-1
unparticle with both vector and axial-vector couplings to leptons. Analytical
expressions were found that agree with previous calculations for the muon MDM in
the limit of flavor conserving couplings.

The muon MDM was numerically evaluated to obtain
bounds on the coupling constants from the experimental measurements assuming
that LFV is mainly dominated by LFV couplings involving the muon and tau
leptons. It was found that the
latest experimental data for the muon MDM favor a scenario in which the spin-1
unparticle contribution is dominated by the vector couplings since the
contribution from axial-vector couplings is negative. It is also found that the decay $\tau\to 3\mu$
can  constrain considerably the allowed magnitude of the $\lambda_V^{\mu\mu}$ and
$\lambda_V^{\mu\tau}$ couplings.

As far as the EDM is concerned, since it depends on several free parameters, we content ourselves with estimating
its order of magnitude. It is found that that the unparticle contributions are well below the experimental
limits of the electron and muon EDMs.

\acknowledgments{We acknowledge support from Conacyt and SNI (M\'exico). Support
from VIEP-BUAP is also acknowledged.}

\appendix
\section{Decay $\dec$}
\label{appendix}
In this appendix we present the calculation for the
$\dec$ decay mediated by spin-1 unparticles, which
proceeds through two tree-level Feynman diagrams (in the first diagram the
unparticle couples to the final $\ell_k^-\ell_k^+$ leptons, while the second
diagram is obtained through the exchange of $\ell_j^-\leftrightarrow \ell_k^-$).
After
introducing the Feynman
rules for the spin-1 unparticle, the decay width can be written as

\begin{equation}
 \Gamma(\dec)=\frac{m_i}{2^9
\pi^3}\left|\frac{A_\du}{\sin (d
\pi)}\right|^2
\left(\frac{m_i}{\Lambda_\uu}\right)^{4 (\du -1)}\int
dx_1\int dx_2\left(|{\cal M}_1|^2+|{\cal M}_2|^2- 2{\rm Re}( {\cal
M}_{12})\right).
\end{equation}
If $j=k$  a factor of $1/(2!)$ must be included as there are two identical
particles in the final state. The integration area on the
$x_1$ vs. $x_2$ plane is

\begin{eqnarray}
4s_k^2-s_j^2 &\le &x_1\le1-2s_j,\nonumber\\
x_2&\gtreqqless&\frac{1}{2}\left(1-x_1\mp\sqrt{\frac{
\left((1-x_1)^2-4s_j^2\right)\left(x_1+s_j^2-4s_k^2\right)} {x_1+s_j^2}
}\right)
\end{eqnarray}
where $s_j=m_j/m_i$.
In addition, ${\cal M}_i$ arises from the squared amplitude
corresponding to each Feynman diagram and ${\cal M}_{12}$ from their interference. These terms depend on the $\du$ dimension and the $x_1$, $x_2$ variables,
and are given by:

\begin{eqnarray}
|{\cal M}_1|^2&=&|\lambda_V^{kk}|^2 \left(
|\lambda_V^{ij}|^2 f_1(s_j,s_k) +  |\lambda_A^{ij}|^2 f_1(-s_j,s_k)\right) +
 |\lambda_A^{kk}|^2 \left(|\lambda_V^{ij}|^2 f_2(s_j,s_k) +  |\lambda_A^{ij}|^2
f_2(-s_j,s_k)\right),
\end{eqnarray}

\begin{eqnarray}
|{\cal M}_2|^2&=& |\lambda_V^{ik}|^2 \left(
|\lambda_V^{jk}|^2 g_1(s_j,s_k) +|\lambda_A^{jk}|^2 g_1(-s_j,s_k)\right)+
 |\lambda_A^{ik}|^2 \left(|\lambda_V^{jk}|^2 g_2(s_j,s_k) + |\lambda_A^{jk}|^2
g_2(-s_j,s_k)\right)\nonumber\\ &+&   2{\rm Re}\left(\left(\lambda_A^{ik}
\lambda_V^{ik}
\lambda_V^{*jk} \lambda_A^{*jk}\right)\right) g_3(s_j,s_k)
\end{eqnarray}

\begin{eqnarray}
{\cal M}_{12}&=& \lambda_A^{ik}
\left(\lambda_A^{kk} \left(
\lambda_V^{*ij}\lambda_V^{*jk}  h_1(s_j,s_k)+\lambda_A^{*ij}\lambda_A^{*jk}  h_1(-s_j,s_k)\right) +
    \lambda_V^{kk} \left(\lambda_V^{*jk}\lambda_A^{*ij}  h_2(s_j,s_k) +
\lambda_A^{*jk}\lambda_V^{*ij}
 h_2(-s_j,s_k)\right)\right)
\nonumber\\&+&\left(V\leftrightarrow A,h_1(s_j,s_k)\leftrightarrow
h_2(-s_j,-s_k)\right),
\end{eqnarray}

For the sake of clarity we omitted the explicit dependence on $x_1$, $x_2$ and
$\du$. The $f_i$, $g_i$ and $h_i$ functions are

\begin{eqnarray}
f_1(u,v)= - \left(2u^3+\left(1+x_1\right)
u^2+2 \left(2v^2+x_1\right)u+\left(x_1-1\right) \left(2
v^2+x_1\right)+2
   \left(x_1+x_2-1\right) x_2\right)\left(u^2+x_1\right)^{2(\du-2)},
\end{eqnarray}

\begin{eqnarray}
 f_2(u,v)&=&f_1(u,-v)+2 v^2 \left(2
   u+x_1-1\right)
   \left(u \left(3
   u+2\right)+2 x_1+1\right)\left(u^2+x_1\right)^{2\du-5},
\end{eqnarray}

\begin{eqnarray}
 g_1(u,v)&=&\frac{1}{2} \Big[u^4 \left(v-1\right) \left(x_2
   \left(v+3\right)+2 v^2+v+1\right)+u^2
   \big(\left(5-x_2\right) v^4-\left(\left(x_2-10\right) x_2+4
   x_1 \left(x_2+1\right)-3\right) v^2\nonumber\\ &+& 2 x_2^2 v-2 v^5+x_2
   \left(-2 x_2^2+x_2-4 x_1 \left(x_2+1\right)+1\right)\big)-
2 x_2^3 \left(2v \left(v+1\right)+2 x_1-1\right)-4
\left(x_1-1\right){}^2\nonumber\\ &-&
2u v \left(2 v+x_2-1\right) \left(v^2+x_2\right)
   \left(v \left(3 v+2\right)+2 x_2+1\right)
   v^4-x_2^2 \left(\left(4 x_1-3\right) v^2+3 v^4+10 v^3+4
   \left(x_1-1\right) x_1\right)\nonumber\\ &-& x_2 v^2 \left(6 v^3+v^2+4
   x_1 \left(2 x_1-3\right)+3\right)-2
x_2^4\Big]\left(v^2+x_2\right)^{2(\du-3)},
\end{eqnarray}

\begin{eqnarray}
 g_2(u,v)=g_1(-u,-v),
\end{eqnarray}

\begin{eqnarray}
 g_3(u,v)=-4  \left(u^2-v^2\right)\left(1+x_2\right)\left(v^2+x_2\right)^{2(\du-2)}
\end{eqnarray}

\begin{eqnarray}
 h_1(u,v)&=&\frac{1}{2}  \Big[u^6
   \left(v+1\right)+u^5 \left(1-x_2(v-1)
   +v^2\right)+u^4 \big(\left(x_1+x_2-4\right)
   v^2+\left(x_1-x_2\right) v\nonumber\\&-&4 v^3+2 \left(x_2^2+x_1
   \left(x_2+1\right)\right)\big)-u^3 \big(-x_2^2
   \left(v+1\right)-x_2 \left(v \left(2
   \left(v-2\right) v-x_1+1\right)+3 x_1\right)\nonumber\\&+&v \left(v
   \left(v \left(4 v+x_1-1\right)-2
   x_1+4\right)+1\right)+x_2-x_1\big)+u^2 \big(-2
   \left(x_1+1\right) v^4+\left(-3 x_1-2 x_2+3\right)
   v^3\nonumber\\&+&\left(3 x_1^2+\left(x_2-7\right) x_1+\left(x_2-7\right)
   x_2+1\right) v^2-x_2 \left(x_1+x_2-1\right) v+2
   \left(x_2-1\right) x_2^2+2 x_1 x_2 \left(3 x_2-1\right)\nonumber\\&+&x_1^2
   \left(4 x_2+1\right)\big)-u \big(-x_1 x_2^2
   \left(v+1\right)-x_1 x_2 \left(v \left(2
   \left(v-2\right) v+1\right)+2 x_1-1\right)\nonumber\\&+&v \left(v
   \left(x_1 \left(v \left(4 v-3\right)+2\right)-v-2
   x_1^2-1\right)+x_1\right)\big)-\left(x_1^2-1\right) v^4+\left(x_1
   \left(1-2 x_2\right)-1\right) v^3\nonumber\\&-&x_1 \left(x_2-1\right) x_2
   v+\left(2 x_2+x_1 \left(2 x_1^2+\left(2 x_2-3\right)
   x_1+\left(x_2-3\right) x_2+1\right)\right) v^2\nonumber\\&+&2 x_1 x_2
   \left(x_1+x_2-1\right) \left(x_1+x_2\right)\Big]
\left(u^2+x_1\right)^{\du-3} \left(v^2+x_2\right)^{\du-3},
\end{eqnarray}

\begin{eqnarray}
 h_2(u,v)&=&\frac{1}{2} \Big[u^3\left(x_2+1\right)
   \left(v-1\right)+u^2 \left(\left(x_1+2 x_2-2\right)
   v^2-(v^2-1)v+x_1+2 x_2 \left(x_1+x_2\right)\right)\nonumber\\&+&u
   \left(\left(2 x_1+3 x_2-1\right) v^3-2 \left(x_1+3
   x_2-1\right) v^2+\left(x_2 \left(2 x_1+3
   x_2-2\right)-1\right) v-6 v^4-x_2 \left(2
   x_1+x_2-1\right)\right)\nonumber\\&+&u^4 \left(v+1\right)+x_2^2 \left(3
   \left(v-1\right) v+4 x_1-2\right)+\left(x_1-1\right) v^2
   \left(v \left(3 v-2\right)+2 x_1-1\right)\nonumber\\&+&x_2 \left(\left(6
   x_1-4\right) v^2+\left(1-2 x_1\right) v-3 v^3+2
   \left(x_1-1\right) x_1\right)+2
x_2^3\Big]\left(u^2+x_1\right)^{\du-2}\left(v^2+x_2\right)^{\du-3}.
\end{eqnarray}

\subsection{Decay $\ell_i\to 3l_j$}

In the  case of this decay, the above expressions simplify considerably.  We can write the decay width
as

\begin{eqnarray}
 \Gamma(\ell_i\to 3\ell_j) &=&\frac{m_i}{2^{10} \pi^3}\left|\frac{A_\du}{\sin (d
\pi)}\right|^2
\left(\frac{m_i}{\Lambda_\uu}\right)^{4 (\du -1)}
\Big[|\lambda_V^{jj}|^2
\left(|\lambda_V^{ij}|^2
\eta_1(s_j,\du)+|\lambda_A^{ij}|^2
\eta_1(-s_j,\du)\right)\nonumber\\&+&
|\lambda_A^{jj}|^2
\left(|\lambda_V^{ij}|^2
\eta_2(s_j,\du)+|\lambda_A^{ij}|^2
\eta_2(-s_j,\du)\right)+2{\rm
Re}\left(\lambda_A^{jj}\lambda_V^{*jj} \lambda_V^{ij}\lambda_A^{*ij}\right)
\eta_3(s_j,\du)\Big],
\end{eqnarray}
with

\begin{equation}
\eta_i(z,\du)=\int dx_1 \int dx_2 \xi_i(z,\du)
\end{equation}
where the integration limits on the $x_1$ vs. $x_2$ plane are now given by
\begin{eqnarray}
3s_j^2 &\le &x_1\le1-2s_j ,\nonumber\\
x_2&\gtreqqless&\frac{1}{2}\left(1-x_1\mp\sqrt{\frac{
\left((1-x_1)^2-4s_j^2\right)\left(x_1-3s_j^2\right)} {x_1+s_j^2} }\right),
\end{eqnarray}
and the $\xi_i$ functions, whose explicit dependence on $x_1$ and $x_2$ has also been omitted for clarity, are given by
\begin{eqnarray}
\xi_1(z,\du)&=&-\frac{1}{2}\Big(2q_1^{2 (\du-2)} \left((z+1) (3 z-1)x_1+x_1^2+2 \left(x_1+x_2-1\right) x_2+(6 z-1)  z^2\right)\nonumber\\&+&
(q_1q_2)^{\du-2} \left(x_1+x_2+z-1\right)
   \left(x_1+x_2+ (3z+1)z\right)\Big)+(x_1\leftrightarrow x_2),
\end{eqnarray}

\begin{eqnarray}
\xi_2(z,\du)&=& \frac{1}{2}\Big(4q_1^{2 \du-5}
   \left(z+1\right)^2 \left(x_1+2 z-1\right)z^2+2q_1^{2 \du-4}
    \left((z+1)^2x_1-2x_2^2-x_1^2-2
   \left(x_1-1\right) x_2-(10 z+3) z^2\right)\nonumber\\&+&
   2q_1^{\du-3} q_2^{\du-2} (z+1) \left(x_1+2 z-1\right)
   \left(x_1+2 z^2+z\right)z\nonumber\\&-&
  \left(q_1
   q_2\right)^{\du-3}z^2 (z+1)^2  \left(x_1x_2+(x_1+x_2) z^2+z \left(z
   \left(z+1\right)-1\right)\right)\nonumber\\&&-\left(q_1
   q_2\right)^{\du-2} \left(\left(x_1+x_2+2\right)
   z^2+x_1^2+x_2^2+2 x_1 x_2-x_1-x_2+(7 z^2-1)z\right)\Big)+(x_1\leftrightarrow x_2),
\end{eqnarray}

\begin{eqnarray}
\xi_3(z,\du)&=&-2\Big(
    q_1^{\du-3} q_2^{\du-2}  (1+ x_1^2-2z^2)z^2
   + q_1^{\du-2} q_2^{\du-2} \left(x_1+x_2-1\right)
   \left(x_1+x_2\right)\Big)+(x_1\leftrightarrow x_2).
\end{eqnarray}
We also introduced the notation $q_i=x_i+z^2$.

\end{document}